\documentclass[aps,pre,showpacs,twocolumn,superscriptaddress,longbibliography]{revtex4-1}
\usepackage[pdftex]{graphicx}
\usepackage{amsmath}
\usepackage[pdftex,bookmarks,colorlinks,breaklinks]{hyperref} % PDF hyperlinks, with coloured links
\hypersetup{linkcolor=red,citecolor=blue,filecolor=dullmagenta,urlcolor=blue} % colored links
\hypersetup{pdftoolbar=true,pdfpagemode=UseNone,pdfstartview=FitH, colorlinks=true,extension=}

\begin{document}
\title{A microwave realization of the chiral orthogonal, unitary, and symplectic ensembles}
\author{A. Rehemanjiang}
\affiliation{Fachbereich Physik der Philipps-Universit\"at Marburg, D-35032 Marburg, Germany}
\author{M. Richter}
\affiliation{School of Mathematical Sciences, University of Nottingham, Nottingham NG7 2RD, UK}
\affiliation{Universit\'{e} C\^{o}te d'Azur, CNRS, Institut de Physique de Nice (InPhyNi), 06108 Nice, France}
\author{U. Kuhl}
\affiliation{Fachbereich Physik der Philipps-Universit\"at Marburg, D-35032 Marburg, Germany}
\affiliation{Universit\'{e} C\^{o}te d'Azur, CNRS, Institut de Physique de Nice (InPhyNi), 06108 Nice, France}
\author{H.-J. St\"ockmann}
\affiliation{Fachbereich Physik der Philipps-Universit\"at Marburg, D-35032 Marburg, Germany}
\date{\today}
\begin{abstract}

Random matrix theory has proven very successful in the understanding of the spectra of chaotic systems.
Depending on symmetry with respect to time reversal and the presence or absence of a spin 1/2 there are three ensembles, the Gaussian orthogonal (GOE), Gaussian unitary (GUE), and Gaussian symplectic (GSE) one.
With a further particle-antiparticle symmetry the chiral variants of these ensembles, the chiral orthogonal, unitary, and symplectic ensembles (the BDI, AIII, and CII in Cartan's notation) appear.
A microwave study of the chiral ensembles is presented using a linear chain of evanescently coupled dielectric cylindrical resonators.
In all cases the predicted repulsion behavior between positive and negative eigenvalues for energies close to zero could be verified.

\end{abstract}


\maketitle

Random matrix theory originally had been developed by Wigner, Dyson, Mehta \cite{dys62c,meh91} and others as a tool to describe the spectral properties of chaotic systems.
For time-reversal symmetric systems the Hamiltonian $H$ commutes with the time-reversal operator $T$, $HT=TH$, where $T^2=1$ for systems without spin 1/2, and $T^2=-1$ in the presence of a spin 1/2 \cite{haa10}.
The three options ($T^2=1$, no $T$, $T^2=-1$) give rise to the three classical random matrix ensembles, the Gaussian orthogonal (GOE), unitary (GUE) and symplectic (GSE) one, respectively.
Further there may be a chiral symmetry, i.\,e., an operator $C$ {\em anti}commuting with $H$, $HC=-CH$, again with the two options $C^2=1$ and $C^2=-1$.
Such a symmetry exists, e.\,g., for the Dirac equation\cite{ver93}.
All possible combinations of ($T^2=1$, no $T$, $T^2=-1$) and ($C^2=1$, no $C$, $C^2=-1$) yield a total of nine random matrix ensembles.
Together with the last remaining option (no $T$, no $C$, but $CT$) one finally ends up with the {\em ten-fold way} \cite{zir96,alt97c}.

For the GOE there is an abundant number of realizations, see Sec.~3.2 of Ref.~\onlinecite{stoe99}, but for the GUE the number of experiments is still small \cite{so95,sto95b,law10}.
The GSE has been realized only recently by us in a peculiarly designed microwave network mimicking a spin 1/2 \cite{reh16}.

For the new ensembles systematic random matrix studies are still missing though there are a lot of studies of systems showing chiral symmetry \cite{bee15}.
In the present work a microwave study of the chiral relatives of the classical ensembles shall be presented, the chiral orthogonal (chOE), the chiral unitary (chUE), and the chiral symplectic (chSE) ensemble (the BDI, the AIII, and the CII in Cartan's notation).
We omit the `G' in the notation, since the ensembles studied by us are partly not Gaussian.

For a chiral symmetry particles and anti-particles are not really needed.
Sufficient is a system consisting of two subsystems I and II with interactions only between I and II, but no internal interactions within I or II.
The Hamiltonian for such a situation may be written as
\begin{equation}\label{eq:chiham}
    H=\left(\begin{array}{cc}
        0 & A \\
        A^\dag & 0
      \end{array}\right)\,,
\end{equation}
where the diagonal blocks belong to the two subsystems, and the off-diagonal blocks describe the interaction.
Hamiltonian (\ref{eq:chiham}) is chiral symmetric, since it anticommutes with $C=\mathrm{diag}({\bf 1}_n,-{\bf 1}_m)$.
The characteristic polynomial of $H$ is given by
\begin{equation}\label{eq:chi}
    \chi(E)= \left|
               \begin{array}{cc}
                 E\cdot{\bf 1}_n & -A \\
                 -A^\dag & E\cdot{\bf 1}_m \\
               \end{array}
             \right|=
    E^{n-m}\left|E^2\cdot{\bf 1}_m-A^\dag A\right|\,,
\end{equation}
where an elementary matrix transformation has been applied.
Equation (\ref{eq:chi}) has a number of important implications:
(i) For $n>m$ there are $\mu=n-m$ zero eigenvalues.
They do not depend on the interaction between the two subsystems and can only be destroyed by lifting the chiral symmetry.
(ii) All other eigenvalues appear in pairs $E_k$ and $E_{-k}=-E_k$ ($k=1,2,\dots$).
(iii) For energies far from $E=0$ the statistical properties of the chiral ensembles approach those of the classical ones, but close to $E=0$ the eigenvalues $E_1, E_2, \dots$ feel the proximity of their partners $E_{-1}, E_{-2}, \dots$, resulting in an oscillatory modulation in the density of states $\rho (E)$ and a possible repulsion of the non-zero eigenvalue pairs close to $E=0$ \cite{iva02},
\begin{equation}\label{eq:rep}
  \rho (E) \sim |E|^{\alpha+\mu\beta}\,,
\end{equation}
where $\beta$ is the universality index known already from the classical ensembles ($\beta=1, 2, 4$ for the GOE, the GUE, and the GSE, respectively), and $\alpha=\beta -1$ for their chiral relatives \cite{bee15}.

\begin{figure} 
	\includegraphics[width=0.85\linewidth]{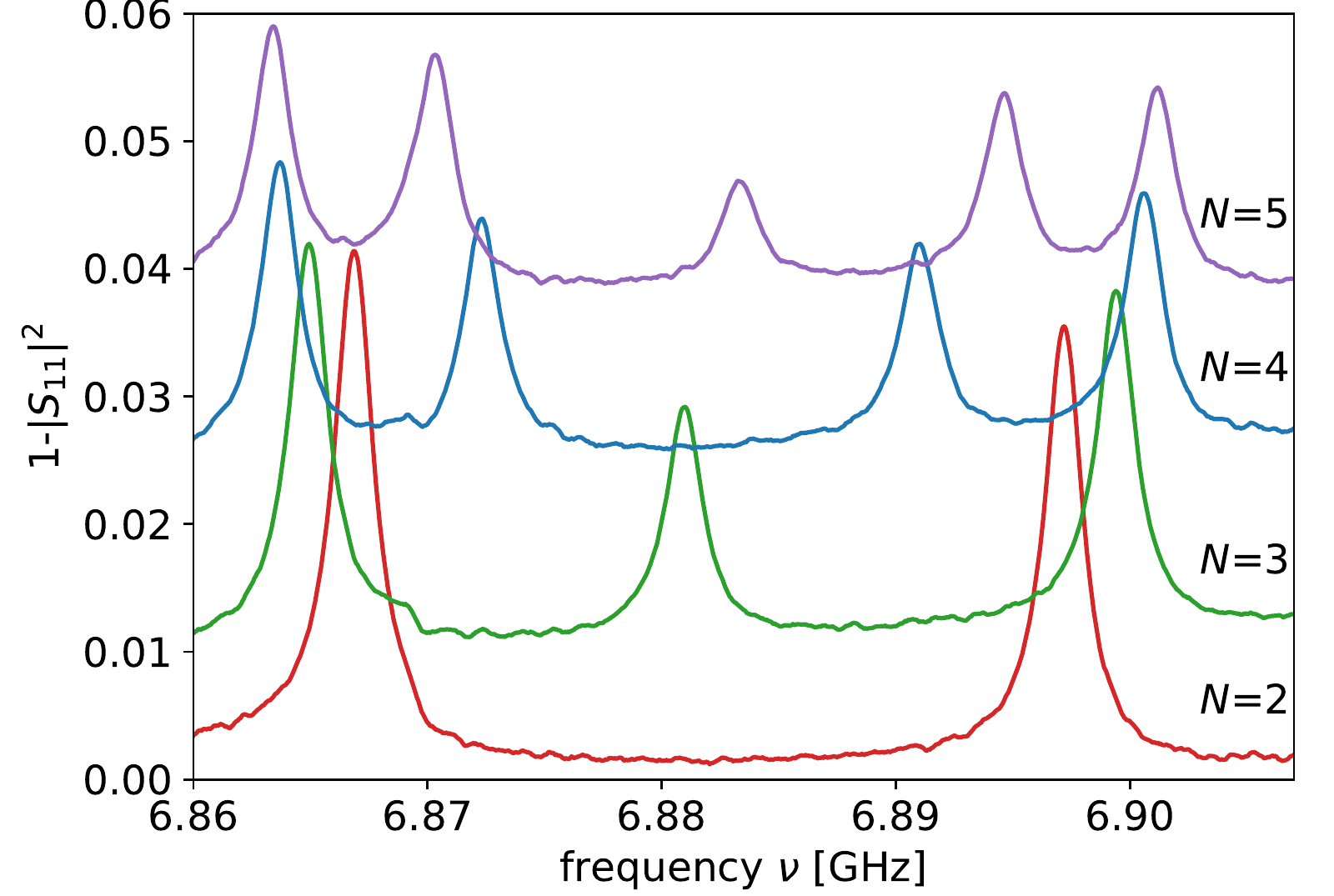}\\
	\includegraphics[width=0.82\linewidth]{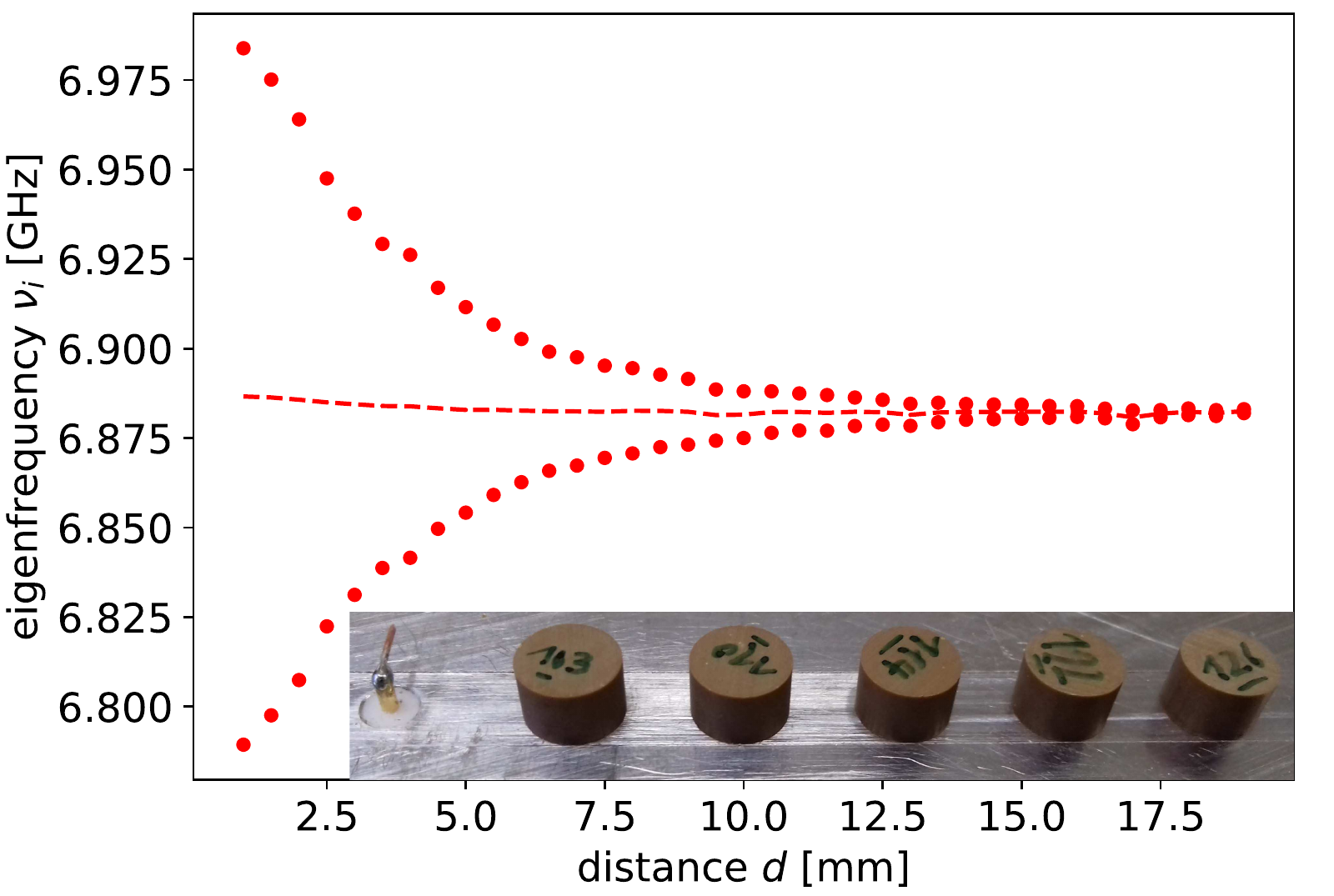}
	\caption{\label{fig:GOE_spectra}
		Top: Spectra for a linear chain with $N$ = 2, 3, 4 and 5 dielectric cylinders.
		Bottom: Eigenfrequencies for a two resonator system in dependence of their
	    distance, used to calibrate
	    the coupling constant $a$ in terms of the distance.
	    The dashed line denotes the center of gravity of the two eigenfrequencies.
	    The inset shows the used set-up.
	    The top plate has been removed for the photograph.
	}
\end{figure}

The main building blocks for our experimental realization of the chiral ensembles are dielectric cylindrical resonators ($h=$5\,mm, $r=$3.8\,mm) with an index of reflection of $n=6$, see inset in Fig.~\ref{fig:GOE_spectra}(bottom).
They are placed between aluminum bottom and top plates separated by a distance of $H$=11\,mm.

There are two types of resonance modes, the transverse magnetic (TM) and transverse electric (TE) one.
In the experiment we used the lowest TE mode with magnetic field parallel and electric field perpendicular to the cylinder axis.
As the resonators were not perfectly equal, the mean eigenfrequency $\nu_0=13.6$\,GHz spread by about 3\,MHz corresponding to 20\,percent of the line width.
In the experiment $\nu_0$ corresponds to the energy zero and $\nu-\nu_0$ to the energy.
As $\nu_0$ is below the cutoff frequency $\nu_\mathrm{cutoff}=c/(2H)=9.369$\,GHz, where $c$ is the velocity of light, the resonators are evanescently coupled allowing to vary the coupling strength via the distance.
For further experimental details it is referred to \cite{bar13a,bel13b}.
Reflections $S_{11}, S_{22}$ and transmissions $S_{12}, S_{21}$ have been measured in the range 6.7 to 7.0\,GHz using a vector network analyzer (Agilent 8720ES).

For the realization of the chOE the set-up shown in
Fig.~\ref{fig:GOE_spectra}(bottom) had been used.
Up to five resonators are placed in a row.
The resonances are excited by a bent antenna placed close to the leftmost resonator.
Figure~\ref{fig:GOE_spectra}(top) shows typical reflection spectra for
$N$ = 2, 3, 4 and 5 resonators obtained by removing one resonator after the other
from the right.
The chiral symmetry is clearly evident from the spectra.
For $N>5$ it became more and more difficult to extract all resonances.
Furthermore localization effects eventually became important.

The situation can be mapped onto a Hamiltonian with the eigenfrequencies of the resonators in the diagonal, the coupling constants in the secondary diagonals, and zeros everywhere else.
Ordering row and columns with odds sites first and even ones secondly, the Hamiltonian takes the structure (\ref{eq:chiham}).
For $N=2$ the eigenfrequencies are given by
$ \nu_\pm=\nu_0 \pm\sqrt{|a|^2+\Delta^2/4}$,
where $\Delta$ is the difference of the eigenfrequencies of the two unperturbed resonators, and $a$ the coupling constant due to the evanescent coupling.
Figure \ref{fig:GOE_spectra}(bottom) shows the eigenfrequencies for a two-resonators system in
dependence of their distance.
Such two-resonator measurements have been used for a calibration of the coupling constants in terms of the
distance.

\begin{figure}
  \includegraphics[width=1.0\linewidth]{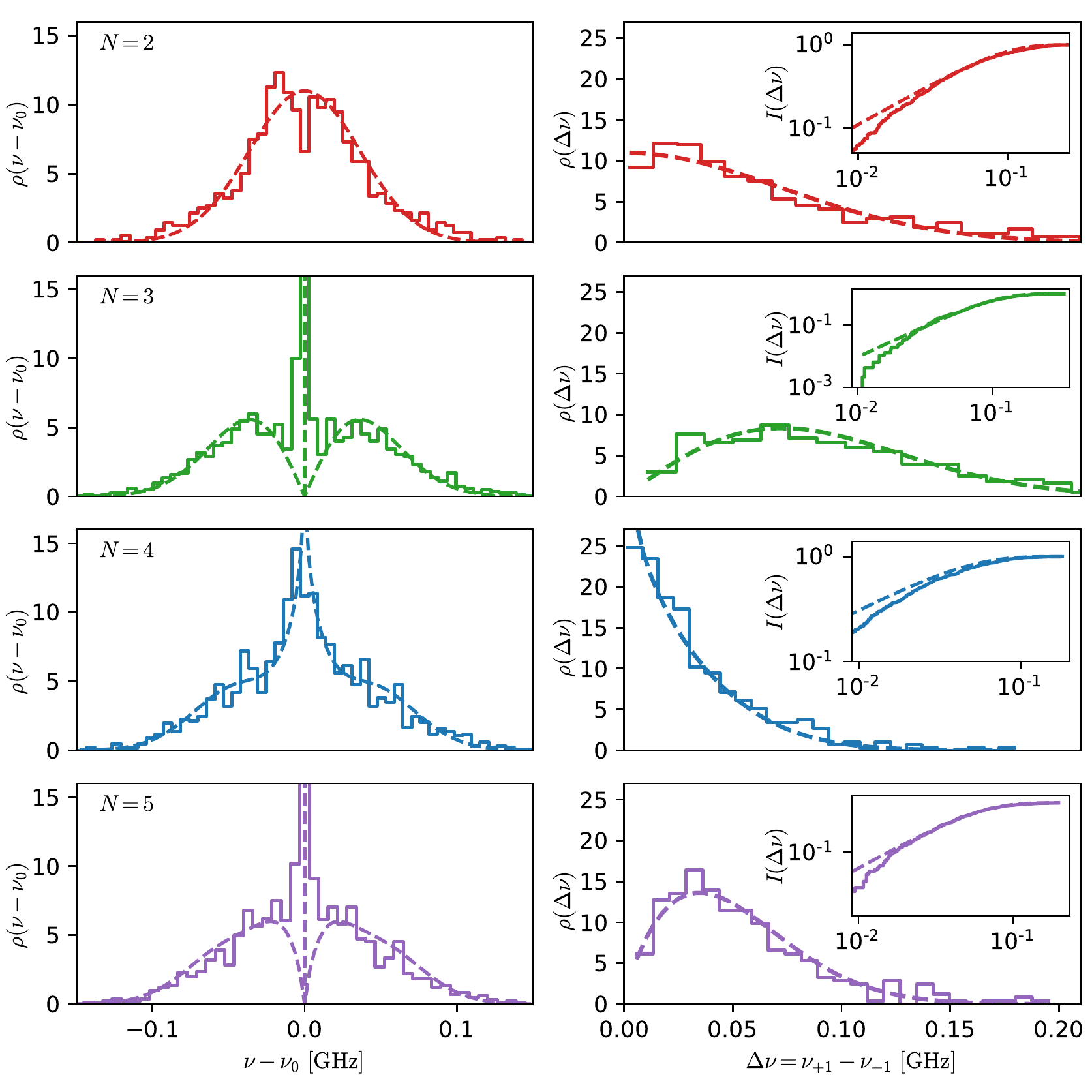}
  \caption{\label{fig:chiGOE}
    Left: Ensemble averaged density of states of the chOE for $N=2$, $3$, $4$, and $5$.
    The dashed lines correspond to the theoretical expectation.
    Right: Two-point correlation function $\rho_{+1,-1}(\Delta\nu)$ for the chOE.
    The dashed lines show the theoretical prediction.
    The inset shows the integrated two-point correlation function $I_{+1,-1}(\Delta\nu)$ in a log-log plot.
  }
\end{figure}

For the chOE some hundred realizations have been taken, with Gaussian distributed coupling constants,
\begin{equation}\label{eq:gauss}
    p_\beta(a)\sim a^{\beta -1}e^{-\frac{a^2}{2\sigma^2}}\,,
\end{equation}
where $\beta=1$, and a cutoff at the maximal available coupling
constant $a_\mathrm{max}= 102.56$\,MHz.
With $\sigma= 0.354\,a_\mathrm{max}$ this means a truncation of 0.5 percent of the area in the tail of the distribution.
For $N$=2 and 3 the A matrix is completely filled, i.e., the resulting ensembles are thus Gaussian.
For increasing $N$ $A$ contains more and more zero matrix elements, hence these ensembles are no longer Gaussian.
While this modifies the details of the modulation of the ensemble averaged density of states close to $E=0$, it does not change the repulsion behavior (\ref{eq:rep}).
Figure~\ref{fig:chiGOE}, left panel, shows the resulting ensemble averaged density of states $\rho(\nu-\nu_0)$ for linear chains of lengths $N=2$, $3$, $4$, and $5$.

For $N\le 5$ the zeros of the characteristic polynomial (\ref{eq:chi}) can be obtained as roots of quadratic equations.
It remains a Gaussian average over the coupling constants to calculate the ensemble averaged density of states \cite{arxRic19}.
The dashed lines in Fig.~\ref{fig:chiGOE}, left panel, show the analytical expressions for $\rho(\nu-\nu_0)$.
For odd $N$ a delta peak at $\nu=\nu_0$ is predicted, and a linear repulsion of the eigenfrequencies, see Eq.~(\ref{eq:rep}).
For even $N$ there should be no repulsion.

All these features are found in the experiment.
There is only one small mismatch resulting from the fact that eigenfrequencies of the resonators are not identical, but differ by some MHz.
Furthermore, the eigenfrequency of the leftmost resonator is detuned by the nearby antenna.
This results in a hole in the distribution at $\nu=\nu_0$ for $N=2$, and to a smaller extent also for $N=4$.

Each non-zero element in the diagonal block of the Hamiltonian
(\ref{eq:chiham}) destroys the chiral symmetry.
This may be due to next-nearest neighbor contributions, or by different
site energies as in the present situation.
This shortcoming is unavoidable: 
Even if the eigenfrequencies of the isolated resonators would be identical they are detuned by the presence of the other ones.
A manifestation of this detuning is the shift of the center of gravity of the two-resonator spectrum observed for small distances, see Fig.~\ref{fig:GOE_spectra}(bottom).

Perturbation of the chiral symmetry results in two imperfections: 
(i) Shift of the center of gravity of the spectrum, 
(ii) A left-right asymmetry between the ``positron'' and the ``electron'' part of the spectrum.
For the linear chain with an odd number of elements this means in particular that the zero energy peak is smeared out.
In addition the repulsion behavior is distorted as is evident, e.\,g., from Fig.~\ref{fig:chiGOE}, left panel, for $N=3$.

Therefore we studied in addition another quantity, the two-point correlation function $\rho_{+1,-1}(\Delta\nu)$ giving the probability density to find a frequency distance $\Delta\nu$ between the states $\nu_1$ and $\nu_{-1}$.
Because of the chiral symmetry $\rho_{+1,-1}(\Delta\nu)$ is, up to a factor of two, identical with the ensemble averaged density $\rho_1(\nu-\nu_0)$ of state 1.
Figure~\ref{fig:chiGOE}, right panel, shows the results, together with the random matrix expectation.
To accentuate the repulsion behavior, the integrated pair correlation function $I_{+1,-1}(\Delta\nu)=\int_0^{\Delta\nu} d\nu'\,\rho_{+1,-1}(\nu')$ is shown in the inset in a log-log plot.
Because of the integration the repulsion exponents increase by one, i.\,e., repulsion exponents should be 1 and 2 for odd and even $N$, respectively.
For all cases the expected behavior is verified.
For odd $N$ the agreement between theory and experiment is nearly perfect.
Only for the lowest shown $\Delta\nu$ values small deviations appear.
Here the number of realizations entering is below ten and the results are no longer statistically trustworthy.
For even $N$ the deviations are somewhat larger, again a manifestation of the fact that the eigenfrequencies of the resonators are not identical.

\begin{figure}
	\includegraphics[width=0.85\linewidth]{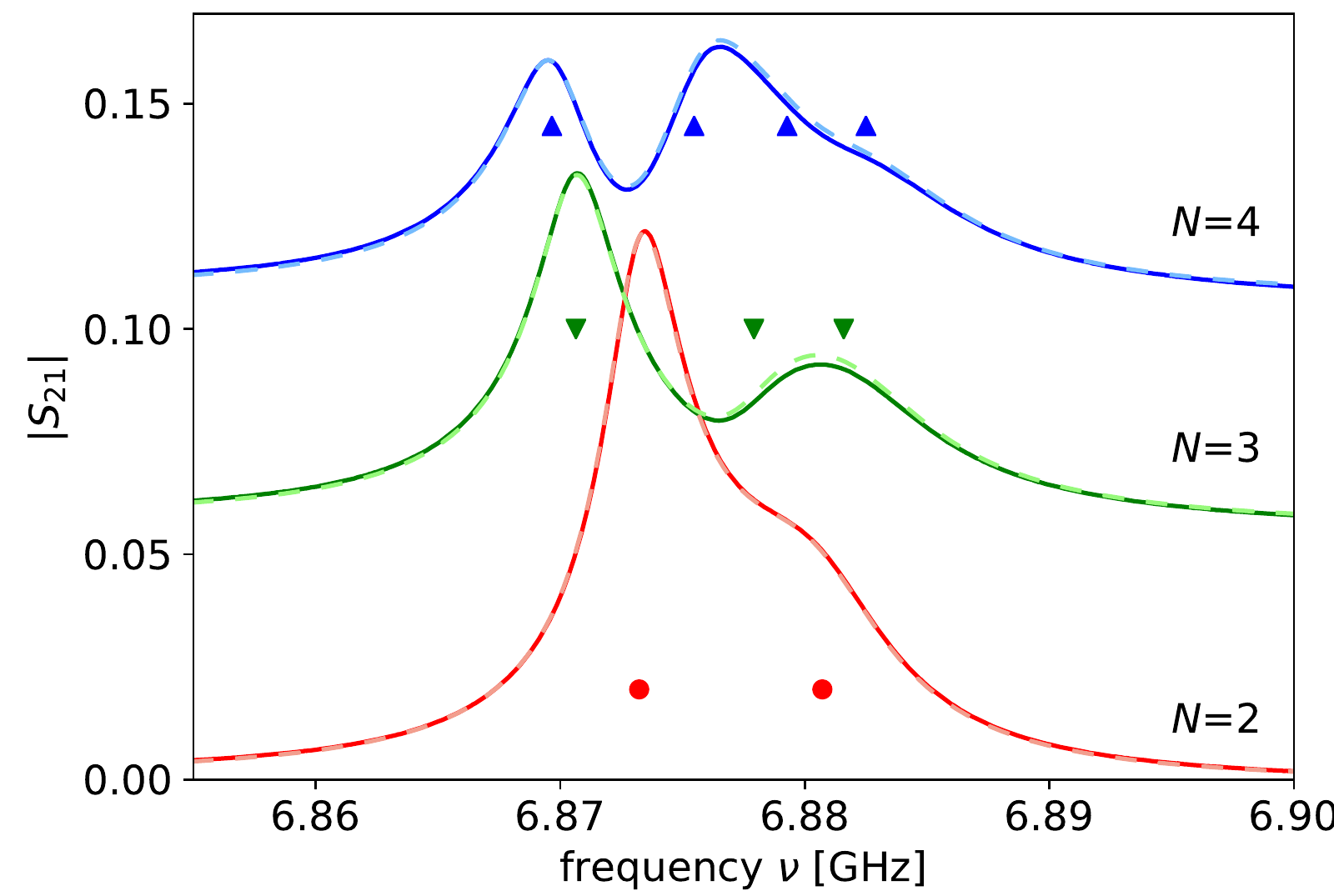}\\
	\includegraphics[width=0.82\linewidth]{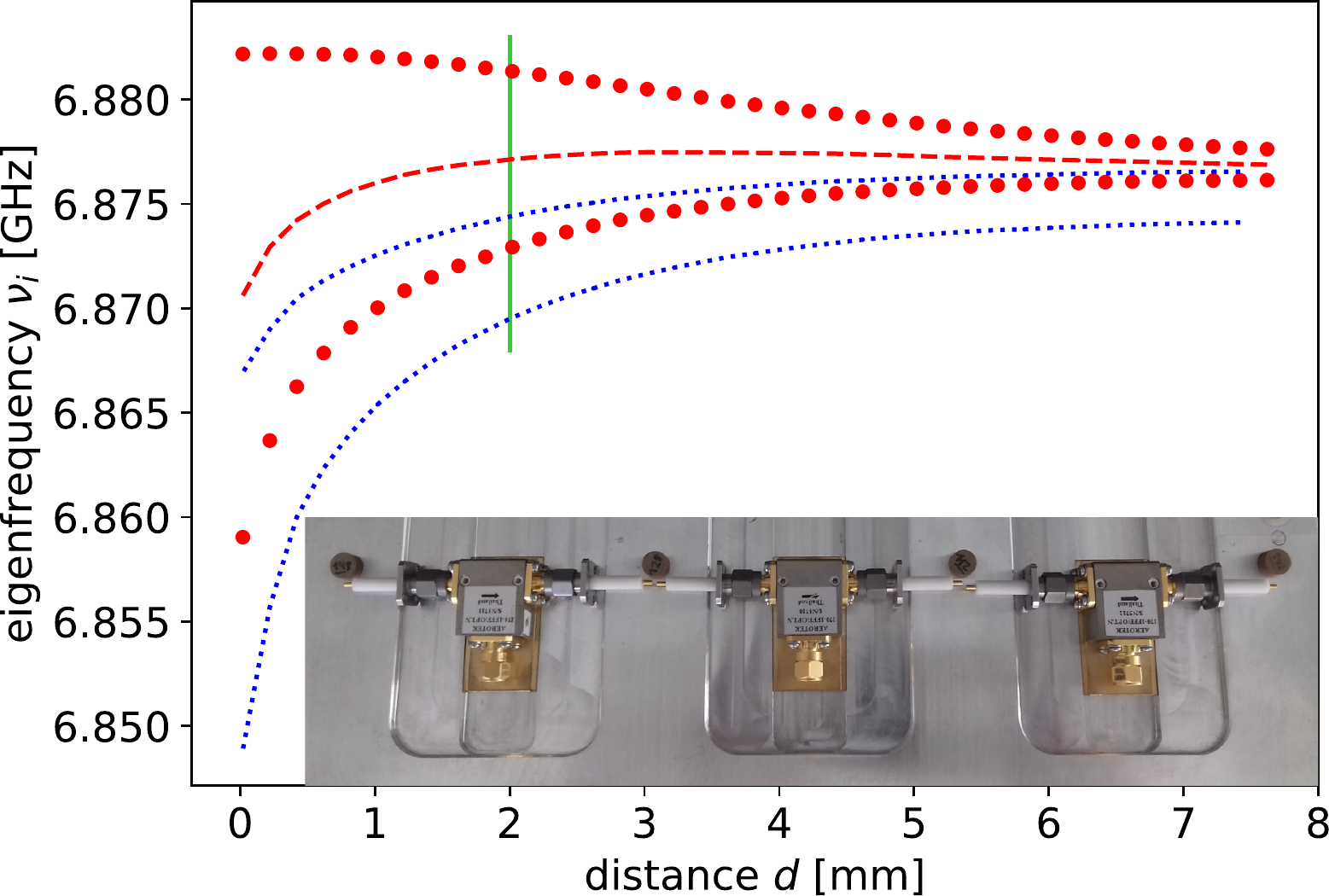}
	\caption{\label{fig:GUE_spectra}
		Top: Reflection spectra for the chUE for $N = 2, 3, 4$ dielectric cylinders.
	    The symbols denote the positions of the resonances extracted by the harmonic inversion technique.
	    The solid lines in dark colors correspond to the measurement, the superimposed dashed lines in light colors to the reconstruction.
		Bottom: Eigenfrequencies for a two resonator system, coupled by a circulator with an open-end (red) and a short-end (blue) side port terminator, respectively, in dependence of the distance.
		The dashed line denotes the center of gravity of the two eigenfrequencies.
		The vertical green line denotes the lower limit of the distances used for the histograms in the left column of Fig.~\ref{fig:chiGUE}.
		The inset shows the used set-up.
	}
\end{figure}

Since the zeros of the characteristic polynomial (\ref{eq:chi}) depend only on the moduli of the coupling constants but not their phases, a convenient way to realize chUE and chSE statistics would be to reuse the same experimental setup and to only take a modified function for the coupling constants, again Eq.~(\ref{eq:gauss}), but now with $\beta=2, 4$ for the chUE and the chSE, respectively.
We preferred a more ambitious approach and studied systems {\em really} having a broken time reversal ($\beta=2$) or symplectic symmetry ($\beta=4$).

In our realization of the chUE the break of time reversal symmetry was achieved by means of circulators, which had been used already for this purpose previously \cite{reh16}.
A circulator is a microwave device with three ports, where waves entering via ports 1, 2, 3 exit through ports 2, 3, 1, respectively.
The inset of Fig.~\ref{fig:GUE_spectra}(bottom) shows the set-up.
Now the resonators are at fixed distances of 85\,mm, too large for a direct coupling which is achieved instead by means of circulators.
The resonances had been excited by loop antennas placed on top of the resonators requiring a height $H=16$\,mm.

Figure~\ref{fig:GUE_spectra}(top) shows typical spectra for $N=2$, $3$, and $4$ resonators.
Unfortunately, the circulators introduce a considerable broadening making an analysis difficult.
Therefore, the harmonic inversion technique has been used, allowing for an analysis of the spectra also for overlapping resonances \cite{wal95,man97a,mai99,kuh08b}.
The deviations between the original and the reconstructed spectra are hardly seen, demonstrating the reliability of the technique.
For $N>4$ a reliable analysis of the spectra was no longer possible.

Since the circulators introduce directionality, the coupling constants are complex.
A Gaussian distribution of the real and imaginary parts results in a distribution $p_2(|a|)$ (see Eq.~\ref{eq:gauss}) for the modulus, and a uniform distribution for the phase.
The phase could be varied by attaching different terminators to the third port of the circulators.
Again a two-resonator measurement was used to relate the distance between circulators and resonators to the coupling.
The red symbols in Fig.~\ref{fig:GUE_spectra}(bottom) show the result for an open-end terminator attached.
The shift of the center of gravity of the two resonances is large, comparable to the splitting.
For short-end terminators (blue dots) the effect is even more dramatic.
Therefore we refrained from varying the phases by attaching different terminators to the circulators and used only the open-end terminators.
But even here for small circulator distances the shift of the center of gravity turned out to be too large.

\begin{figure}
	\includegraphics[width=1.0\columnwidth]{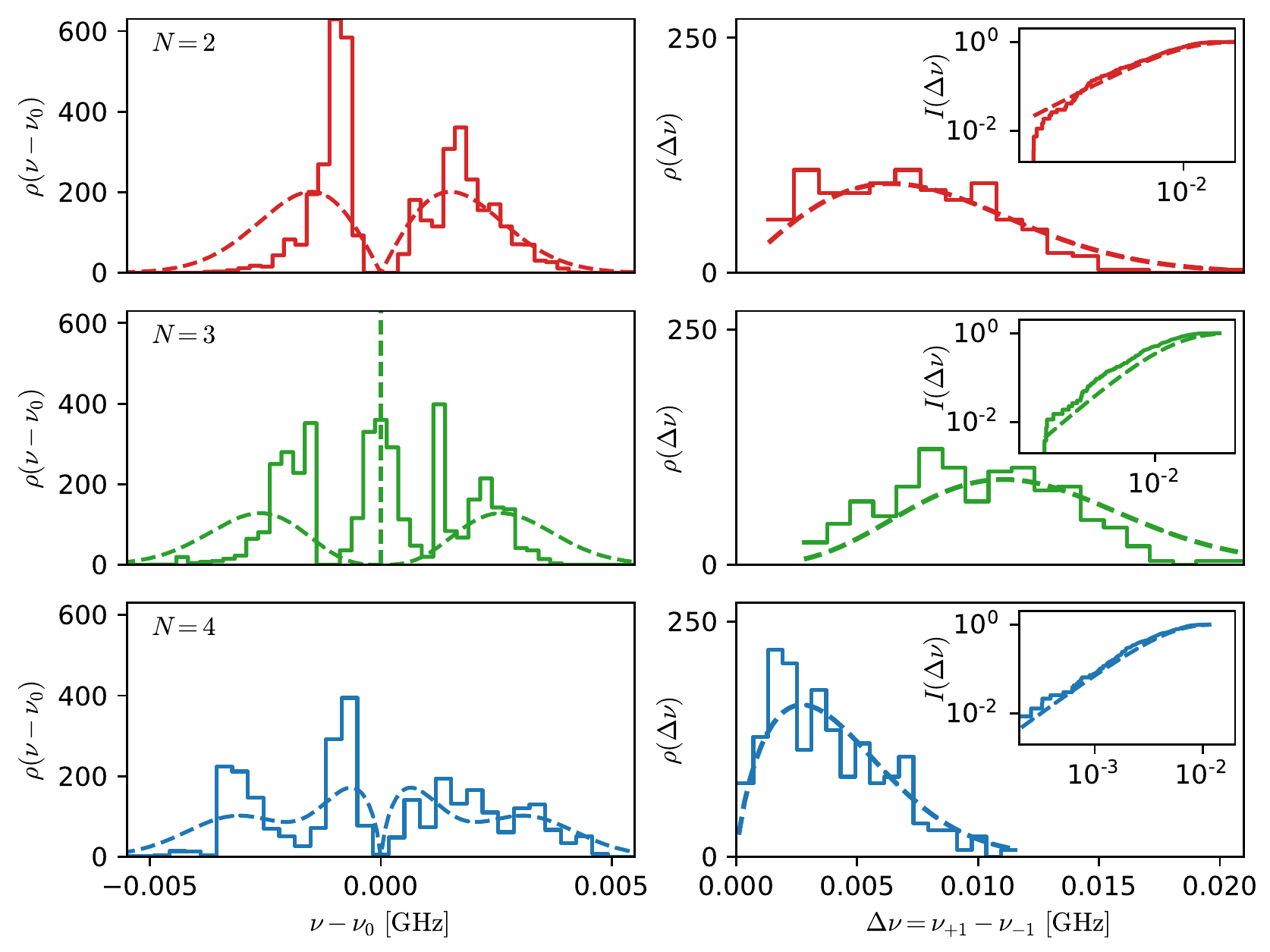}\\
	\caption{\label{fig:chiGUE}
	As Fig.~\ref{fig:chiGOE}, but for the chUE.
	}
\end{figure}

Therefore, only distances $d\ge 2$\,mm were used corresponding to $a_\mathrm{max}= 4.47$\,MHz, with $\sigma= 0.288\,a_\mathrm{max}$.
The results are shown in the left column of Fig.~\ref{fig:chiGUE}.
Though there are significant deviations in detail, the general features of the theoretical predictions are correctly reproduced.
For $N=3$ and $4$ the center of gravity of the spectra had been adjusted individually by less than 1\,MHz.
This shift is caused by a detuning of the eigenfrequencies of the inner resonators due to the presence of the second circulators, an effect not accounted for by the calibration.
All these deficiencies drop out for the two-point correlation function $\rho_{+1,-1}(\Delta\nu)$.
Therefore, for this quantity the whole $d$ range depicted in Fig.~\ref{fig:GUE_spectra} could be used.
The results are presented in the right column of Fig.~\ref{fig:chiGUE}.
The solid lines correspond to the theoretical expectations for the unperturbed systems.
The inset, showing the integrated two-point correlation in a log-log plot, illustrates that for the chUE, too, the expected repulsion exponents are reproduced.

\begin{figure}
	\includegraphics[width=.49\columnwidth]{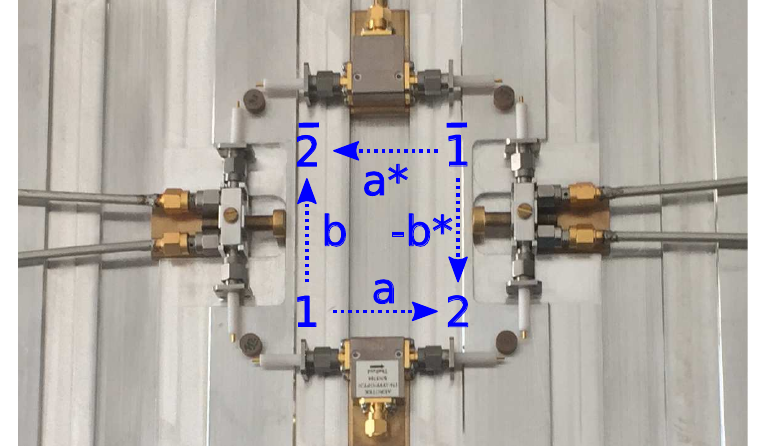}
	\includegraphics[width=.49\columnwidth]{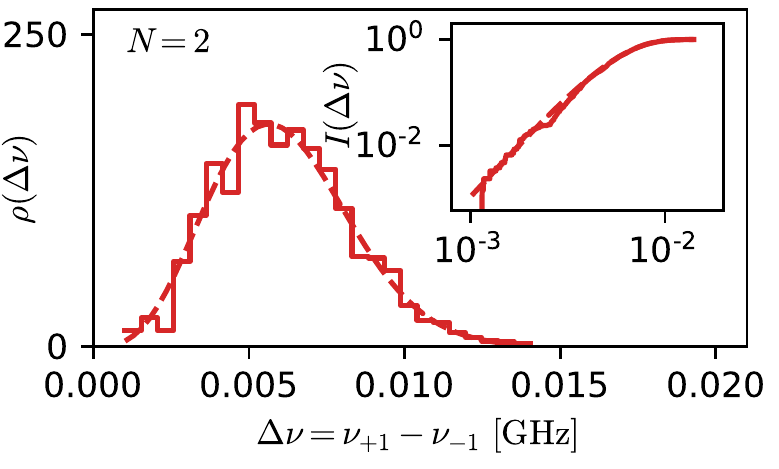}
	\caption{\label{fig:4disks}
	    Left: Set-up to realize the chSE.
	    Right: As Fig.~\ref{fig:chiGOE}, right column, but for the chSE.
	}
\end{figure}

Recently we realized the Gaussian symplectic ensemble in a microwave graph mimicking a spin 1/2 \cite{reh16}.
The main ingredients had been two subgraphs, complex conjugates of each other, coupled by a pair of bonds with a phase shift of $\pi$ in one of the bonds and no phase shift in the other one.
These ideas may be taken over one-to-one to the present situation.
The minimal configuration needs four resonators, $1$, $\bar{1}$, $2$, $\bar{2}$, see Fig.~\ref{fig:4disks}(left).
Resonators $1$ and $2$, and resonators $\bar{1}$ and $\bar{2}$ are coupled by circulators with opposite sense of propagation thus making the two subsystems complex conjugates of each other.
Resonators $1$ and $\bar{2}$, and resonators $\bar{1}$ and $2$ are coupled by cables with a length difference of 1.762\,cm corresponding to a phase difference of $\pi$ at $\nu=\nu_0$.

Though four resonators are involved, the configuration corresponds to $N=2$, where the resonators with and without bars mimic the two spin components.
The corresponding characteristic polynomial is given by
\begin{equation}\label{}
  \chi(E)=|E\cdot{\bf 1} -H|= \left(E^2-|a|^2-|b|^2\right)^2\,.
\end{equation}
There are hence doubly degenerate eigenvalues at
\begin{equation}\label{eq:E}
  E_{+1,-1}=\pm\sqrt{|a|^2+|b|^2}\,.
\end{equation}
The four-resonator system thus shows both the chiral symmetry, with $E_{+1}$ and $E_{-1}=-E_{+1}$ coming in pairs, and the symplectic symmetry, with the characteristic Kramers doublet structure of the spectrum.

Again the harmonic inversion technique was used to analyze the spectra.
Figure~\ref{fig:4disks}(right) shows the distance distribution of the two Kramers doublets, where the distribution function $p_2(|a|)$ (see Eq.~\ref{eq:gauss}) was used both for $|a|$ and $|b|$ (with $a_\mathrm{max} = 7.13$\,MHz and $\sigma = 0.322\,a_\mathrm{max}$).
Hence, Eq.~(\ref{eq:E}) implies a cubic repulsion of $E_{+1}$ and $E_{-1}$, in accordance with Eq.~(\ref{fig:GUE_spectra}).
This repulsion is perfectly reproduced in the experiment.

An extension to larger $N$ values seem hardly feasible.
Each additional pair of resonators would mean four more bonds and a corresponding increase of absorption.
Therefore we have to be content with this demonstration for $N=2$.

\begin{acknowledgments}
H.~Schomerus, Lancaster University, is thanked for illuminating discussions.
This work was funded by the Deutsche Forschungsgemeinschaft via the individual grants STO 157/16-2 and KU 1525/3-1 as well as the European Commission through the H2020 programm by the Open Future Emerging Technology ``NEMF21'' Project (664828).
\end{acknowledgments}

\end{document}